\documentclass[a4paper,11pt]{article}

\usepackage{aas_macros,amsmath,amssymb,comment,cite,esint,graphicx,mathtools}
\usepackage[margin=.8in,letterpaper]{geometry}
\usepackage[colorlinks=true]{hyperref}
\usepackage[affil-it]{authblk}
\usepackage{subcaption}
\usepackage[utf8]{inputenc}
\usepackage{mathrsfs}
\usepackage{appendix}
\usepackage{amssymb}
\usepackage{float}
\usepackage{color}
\usepackage{xcolor} 
\usepackage{cite}
\usepackage{hyperref}
\hypersetup{pageanchor=false}
\usepackage{indentfirst}
\usepackage{url}
\usepackage{float}
\usepackage{caption}
\usepackage[numbers,square,comma,sort&compress,merge]{natbib}
\usepackage[section]{placeins}
\usepackage{esint}
\usepackage{overpic}
\usepackage{graphicx}
\usepackage{epsf,amsmath,bbold,amsfonts,stmaryrd}
\usepackage{textcomp}
\usepackage{ulem}
\usepackage{tikz}

\numberwithin{equation}{section}
\let\originalleft\left
\let\originalright\right
\renewcommand{\left}{\mathopen{}\mathclose\bgroup\originalleft}
\renewcommand{\right}{\aftergroup\egroup\originalright}
\mathcode`\*="8000
{\catcode`\*=\active\gdef*{\mathclose{}\,\mathopen{}}}

\newcommand{\td}[1]{\tilde{#1}}

\newcommand{\be}{\begin{equation}}
\newcommand{\ee}{\end{equation}}
\newcommand{\bea}{\setlength\arraycolsep{2pt} \begin{eqnarray}}
\newcommand{\eea}{\end{eqnarray}}
\newcommand{\nn}{\nonumber}

\newcommand{\beqn}{\begin{eqnarray}}
\newcommand{\eeqn}{\end{eqnarray}}

\def\a{\alpha}
\def\b{\beta}

\def\D{\Delta}
\def\f{\frac}
\def\g{\gamma}

\def\m{\mu} 
\def\n{\nu} 
\def\nn{\nonumber}
\def\pl{\partial}
\def\p{\phi} 
\def\td{\tilde} 

\def\s{\sigma}

\def\t{\theta}

\def\ep{\epsilon}
\def\o{\omega}

\def\be{\begin{equation}}
\def\ee{\end{equation}}
\def\bag{\begin{aligned}}
\def\eag{\end{aligned}}
\def\bea{\begin{eqnarray}}
\def\eea{\end{eqnarray}}
\def\ba{\begin{array}}
\def\ea{\end{array}}

\def\bc{\begin{center}}
\def\ec{\end{center}}

\begin{document}
\title{Extracting energy from plunging region of a Kerr-Taub-NUT black hole by magnetic reconnection}
\author{Zhengwei Cheng$^{1}$,  Songbai Chen$^{1,2}$\footnote{Corresponding author: csb3752@hunnu.edu.cn},
Jiliang Jing$^{1,2}$ \footnote{jljing@hunnu.edu.cn}}

\date{}

\maketitle
\vspace{-10mm}

\begin{center}
{\it	
    $^1$Department of Physics, Institute of Interdisciplinary Studies, Hunan Research Center of the Basic Discipline for Quantum Effects and Quantum Technologies, Key Laboratory of Low Dimensional Quantum Structures
    and Quantum Control of Ministry of Education, Synergetic Innovation Center for Quantum Effects and Applications, Hunan
    Normal University,  Changsha, Hunan 410081, People's Republic of China
    \\
    $ ^2$Center for Gravitation and Cosmology, College of Physical Science and Technology, Yangzhou University, Yangzhou 225009, People's Republic of China
}
\end{center}

\vspace{8mm}

\begin{abstract}
	\vspace{4mm}
 We have studied the energy extraction from a Kerr-Taub-NUT black hole via magnetic reconnection occurring in the plunging region. Our results show that the gravitomagnetic charge suppresses the energy extraction process through magnetic reconnection and reduces the corresponding extraction efficiency, which is opposite to the effects of the black hole spin and the magnetization parameter. Finally, we treat the energy extraction process through magnetic reconnection as a mechanism to revisit the problem of the observed jet power and radiative efficiency of GRS 1915+105. Our results show that the allowed black hole parameter region originating from the jet power has an intersection with the region from the radiative efficiency. This means that with this mechanism related to magnetic reconnection the Kerr-Taub-NUT metric can simultaneously explain the observed jet power and radiative efficiency for GRS 1915+105, which is not explained by other mechanisms in previous studies.  
\end{abstract}

\maketitle

\newpage
\baselineskip 18pt

\section{Introduction}\label{sec1} 

Extracting energy from rotating black holes has been an intriguing topic since it was proposed by Penrose.  In the Penrose process, a particle splits into two parts in the ergosphere. One of the parts, possessing negative energy viewed from infinity,   falls into the black hole. The resultant part escapes from the ergosphere to infinity with more energy than the original one because of the law of energy conservation. This implies that one can extract the rotational energy of a Kerr black hole through this process, until the black hole’s mass is reduced to its irreducible value \cite{Christodoulou:1970wf}. However, initiation mechanisms are lacking in the original Penrose process \cite{Wald:1974kya}. One plausible mechanism is the superradiance of a massive bosonic field because a superradiant cloud can be formed with a characteristic spectrum surrounding the black hole \cite{Brito:2015oca}. 
The Blandford-Znajek process \cite{Blandford:1977ds} is another mechanism for extracting the energy of a rotating black hole,  where the magnetic field lines originating from accretion flows near the black hole are assumed to intersect the event horizon and possess an angular velocity smaller than that of the horizon. 
 The Blandford-Znajek (BZ) mechanism has been extensively investigated since it is expected as a promising engine for generating relativistic jets \cite{Komissarov:2005wj, Parfrey:2018dnc, Lu:2023bbn}. 

 Another energy extraction mechanism has been investigated over the past twenty years, which is based on magnetic reconnections in plasmoids \cite{yamada2009, SP1, SP2, Petschek, rela-SP,RevModPhys.82.603,Fan:2024fcy,Lin_2021,Shen:2024wny,Lazarian_2014,Landinez:2024cyu,anti-Petschek1,anti-Petschek2,rela-SP,substorm,Aimar:2023kzj,Ripperda:2020bpz}. 
 Magnetic reconnection refers to the sudden rearrangement of magnetic fields induced  by the local interaction of magnetic field lines with opposing polarities. Such topological changes of magnetic fields  yield a rapid release of magnetic energy into thermal and kinetic energy of plasma outflows. One of the famous classical magnetic reconnection models is the Sweet-Parker model \cite{SP1, SP2}. In this model, magnetic reconnection proceeds with the formation of an elongated diffusion region. However, its predicted reconnection rate is excessively low, failing to account for the observed phenomena in most astrophysical systems. To address this challenge, Petschek \cite{Petschek} proposed a fast magnetic reconnection model in which  the diffusion region is localized and the outflow is bounded by a pair of standing slow mode shocks. Within this reconnection geometry,  the thickness and length of the diffusion region reside on the microscopic scale, resulting in a large aspect ratio for the diffusion region. This, in turn, produces a faster reconnection rate that is sufficient to account for the observed time scales of solar flares and geomagnetic substorms. Moreover, the Petschek model means that slow mode shocks could play an important role in reconnection. However, numerical simulations have shown that Petschek reconnection cannot form in two-dimensional resistive-MHD simulations if the resistivity is uniform \cite{1986Magnetic,Uzdensky2000Two}. To  further understand fast magnetic reconnection, a series of theoretical investigations and numerical simulations have been performed both in collisionless  \cite{2001JGR...106.3737B} and collisional plasmas \cite{Kowal_2009,Loureiro_2007}, and different mechanisms are analyzed from the perspectives of plasma instabilities \cite{Sironi:2022hnw,French:2022zfv,Dong:2022crn,Comisso:2023ygd}, anomalous resistivity\cite{Che_2017,Malyshkin_2005,Singh:2007sn}, turbulence \cite{1999ApJ...517..700L,Li_2023,Franci_2022,Medina_Torrej_n_2021,Price_2016,Kobak_2000} and Hall effects \cite{Malyshkin:2008he,Uzdensky:2006mb,Pucci_2017,Papini_2019,Andr_s_2016}.  Recently, based on magnetic reconnection, Koide and Arai first discussed the possibility of extracting energy from black holes \cite{Koide:2008xr}.  The fast reconnection model has also been applied to study the energy extraction through the plasmoids around black holes \cite{Comisso:2020ykg,Liu2017},  which shows that the energy extraction rate could be comparable to that of the BZ process. Due to its astrophysical significance, the energy extraction by magnetic reconnections has been studied in different types of spacetimes \cite{Khodadi:2022dff, Carleo:2022qlv, Wei:2022jbi, Liu:2022qnr, Wang:2022qmg, Li:2023nmy, Li:2023htz, Ye:2023xyv, Khodadi:2023juk, Shaymatov:2023dtt, Zhang:2024rvk,Zeng:2025vjt,Long:2024tws}.  However, most of the above studies on energy extraction via magnetic reconnection have been mainly focused on circular accretion flows outside the ISCO.  This is because the circular orbits within the ISCO are unstable and the accretion flows inside the ISCO  plunge toward the event horizon with spiral-shape orbits. Recently, energy extraction from a Kerr black hole has been studied through magnetic reconnection within the plunging region \cite{Chen:2024ggq}. Since the plunging region is near the black hole event horizon, the behavior of the ejected plasmoids produced from magnetic reconnection and the corresponding energy extraction carry significant features related to the background spacetime, which could provide a potential way to test theories of gravity. Therefore, it is necessary to further study energy extraction by magnetic reconnections from the plunging region in other background spacetimes of rotating black holes.

The Kerr-Taub-NUT (KTN) metric is an important stationary and axisymmetric metric, which satisfies Einstein field equations with gravitomagnetic monopole and dipole moments \cite{10.2307/1969567,Demianski-Newman1966,Carter1966,Kinnersley1969,Kramer-Neugebauer1968,Robinson-etal1969,Talbot-1969}. It owns three parameters: the mass $M$, the spin parameter $a$ and the NUT charge $l$.  The charge $l$ plays the role of a magnetic mass inducing a topology in the Euclidean section.  Moreover, due to the existence of the NUT charge, KTN spacetime is asymptotically non-flat and there exist conical singularities on the axis of symmetry \cite{Bonnor1969}.  According to Misner  \cite{Misner:1963fr}, this singularity can be avoided by introducing a periodic time coordinate. However, the periodicity condition inevitably gives rise to closed timelike curves (CTCs) in this spacetime, thereby leading to violations of causality and the incompleteness of geodesics \cite{Hawking:1973uf}. Fortunately, these problems can be resolved by giving up the time periodicity condition as well as interpreting the conical singularities as “a linear source of pure angular momentum” \cite{Bonnor1969,Miller1972,Dowker1974,Ramaswamy-Sen1981}.  Although the KTN spacetime exhibits certain unfavorable physical characteristics, it remains highly attractive for exploring various physical phenomena in general relativity. The black hole shadow is found to increase with the NUT charge \cite{Abdujabbarov:2012bn}. Moreover, the NUT charge affects  particle accelerations in the spacetime and further modifies the constraints on the rotation parameter $a$ when the arbitrarily high center-of-mass energy appears in the collision of two particles \cite{Liu:2010ja}.  The effects of the NUT charge on the inner-most stable circular orbits and the perfect fluid disk around the black hole have also been studied in Refs.\cite{Chakraborty:2013kza,Garcia-Reyes:2004mpa}, respectively. The parameter constraint on the Kerr-Taub-NUT black hole is carried out with the observed jet power and radiative efficiency \cite{PhysRevD.108.103013}. The observational possibilities of NUT charge are also analyzed in terms of the shifting of spectral lines from quasars, supernovae and active galactic nuclei \cite{RevModPhys.70.427}. These results may help to probe the gravitomagnetic masses in astronomical observation in the future. 
The main purpose of this paper is to study energy extraction via magnetic reconnection in the plunging region near a Kerr-Taub-NUT black hole and \textcolor{blue}{to} probe corresponding effects from NUT charge.

The paper is organized as follows. In Sec.~\ref{sec:ktn}, we briefly introduce the Kerr-Taub-NUT spacetime and magnetic reconnection. In Sec.~\ref{sec:EE_inspiral}, we analyze energy extraction by magnetic reconnection in the Kerr-Taub-NUT spacetime and probe how the NUT charge affects the corresponding energy extraction. In Sec.~\ref{sec:ExpGRS1915}, we treat the energy extraction process through magnetic reconnection as a mechanism to revisit the problem on the observed jet power and radiative efficiency of GRS 1915+105. Finally, we conclude with a summary.

\section{The magnetic reconnection  in the Kerr-Taub-NUT spacetime}
\label{sec:ktn}
The Kerr-Taub-NUT metric describes the gravitational field of a black hole with the mass parameter $M$, the gravitomagnetic charge $l$, and the rotational parameter $a$. Its form reads~\cite{10.2307/1969567,Demianski-Newman1966,Carter1966,Kinnersley1969,Kramer-Neugebauer1968,Robinson-etal1969,Talbot-1969}
\begin{eqnarray}
\nn
ds^2&=&-\frac{1}{\Sigma}(\Delta-a^2 \sin^2\theta )dt^2+ \frac{\Sigma}{\Delta}dr^2+\Sigma d\theta^2+\frac{1}{\Sigma}\left[ (\Sigma+a\chi)^2\sin^2\theta-\chi^2\Delta \right]d\phi^2 \nonumber\\ 
&+&\frac{2}{\Sigma}\left[ \Delta \chi -a(\Sigma+a\chi)\sin^2\theta\right]d\phi dt \ ,
\end{eqnarray}
where $\Delta$, $\Sigma$, and $\chi$ are defined as 
\begin{eqnarray}
\Delta&=&r^2+a^2-l^2-2Mr \ ,\quad\quad\quad
\Sigma=r^2+(l+a\cos\theta)^2 \ ,\quad\quad\quad
\chi=a\sin^2\theta-2l\cos\theta \ .
\end{eqnarray}
By setting $\Delta=0$, one can obtain the radius of the outer horizon for the black hole
\begin{equation}\label{eh}
    r_H=M+\sqrt{M^2-a^2+l^2} .
\end{equation}
Correspondingly, the radius of the outer  ergoregion surface is given by
\begin{eqnarray}
r_e=M+\sqrt{M^2-a^2 \cos^2\theta+l^2}.
\end{eqnarray}
At the two poles ($\theta=0$ or $\pi$), the outer ergoregion surface touches the event horizon. On the equatorial plane ($\theta=\pi/2$), however, the ergoregion radius $r_e|_{\theta=\frac{\pi}{2}}=M+\sqrt{M^2+l^2}$ depends only on the NUT parameter and the mass, and  has no connection to the black hole spin. By setting the denominator of the metric component $g_{tt}$ to zero, we obtain the location of the singularity 
$$r=0,\quad\quad \text{and}\quad\quad \theta=\cos^{-1}(-l/a)\, .$$ 
When $l>a$, the spacetime becomes free of singularities. The Kerr-Taub-NUT spacetime describes a black hole when $|a| \leq \sqrt{M^2+l^2}$ and a naked singularity when $a > \sqrt{M^2+l^2}$. 

Let us now probe magnetic reconnection in the Kerr-Taub-NUT spacetime. As in \cite{Chen:2024ggq}, we adopt the perfect fluid approximation for the ejected plasmoids and the bulk plasma in the plunging region near the black hole. In this approximation, the stress-energy tensor of the plasma is
\be
    T^{\m\n} =\o\, U^{\m}U^{\n} + p\, g^{\m\n},
\ee
where $p$ and $\o$ are the fluid's proper pressure and enthalpy density, while $U^{\m}$ is the 4-velocity of the streamlines of the fluid. With the $3+1$ form of a Kerr-Taub-NUT metric, 
\bea
    ds^2 = g_{\m\n} dx^{\m} dx^{\n} = -\a^2 dt^2 + \sum_{i = r, \t, \p}  \big(h_i dx^i - \a \b^i dt \big)^2 \, ,
\eea
the energy conservation law of the fluid, $\nabla_{\m} T^{\m}_{\,\,\,\,t} = 0$, can be written as 
\bea
    \pl_t e + \f{1}{h_1h_2h_3} \pl_i ( h_1h_2h_3 S^i) = 0 \, ,
    \label{energy}
\eea
where  
\be\label{alpha}
    \a = \sqrt{-g_{tt} + \f{g^2_{t\p}}{g_{\p\p}}} \, ,\quad  h_i = \sqrt{g_{ii}} \, , \quad  
    \b^i = \delta_{i \p}\f{\sqrt{g_{\p\p}}\, \o^{\p}}{\a} \,, 
\ee
and  $\o^{\p} = -g_{t\p}/g_{\p\p}$ is the angular velocity of frame dragging. The quantity $e =- \a T^t_{\, \, \, t}$ is the energy density at infinity and $S^i = -\a T^i_{\, \, \, t}$ is the flux density. 

For the plasma moving on the equatorial plane, the flow is along timelike geodesics without $\theta$-dependence. Therefore, one has $U_\m dx^\m= -E dt +U_r(r) dr + L d\p$, where $E$ and $L$ \textcolor{blue}{are} the conserved energy and angular momentum along the streamline. To properly describe the  physical processes within the fluid during magnetic reconnection, one must project the fluid 4-velocity into the normal tetrad of the  zero-angular momentum observers (ZAMOs) 
\bea \hat{U}^{(a)} = U^\m \hat{e}_{\, \m}^{(a)} = \hat{\g}_s\left\{1,\hat{v}^{(r)}_s,0,\hat{v}^{(\p)}_s \right\} = \left\{ \f{E - \o^{\phi}L}{\a}, \sqrt{g_{rr}} \, U^r, 0, \, \f{L}{\sqrt{g_{\phi\phi}}} \right\} \, ,
\eea
where the normal tetrad $e^{\mu}_{(a)}$ are
\bea
\hat{e}^{\, \m}_{(t)} = \f{1}{\a} (\pl_t^{\, \m} + \o^{\p} \pl_{\p}^{\, \m}) \, ,\quad\quad  \hat{e}^{\, \m}_{(r)} = \f{1}{h_r} \pl_r^{\, \m} \, ,\quad \quad    \hat{e}^{\, \m}_{(\t)} = \f{1}{h_\t} \pl_\t^{\, \m} \, ,\quad \quad  \hat{e}^{\, \m}_{(\p)} = \f{1}{h_{\p}} \pl_{\p}^{\, \m} \,.
\eea 
With the normal tetrad $e^{\mu}_{(a)}$, the tetrad for the fluid’s rest frame can be defined as
\bea
    &&e_{[0]} = \hat{\g}_s \left[\hat{e}_{(t)}+\hat{v}^{(r)}_s \hat{e}_{(r)}+\hat{v}^{(\p)}_s \hat{e}_{(\p)}\right]  \, , \quad\quad\quad 
    e_{[1]} = \f{1}{\hat{v}_s} \left[\hat{v}^{(\p)}_s \hat{e}_{(r)}-\hat{v}^{(r)}_s \hat{e}_{(\p)}\right]  \, , \nn\\
     &&e_{[2]}  = \hat{e}_{(\t)}  \, , \quad\quad\quad 
   e_{[3]} = \hat{\g}_s \left[\hat{v}_s \hat{e}_{(t)}+\f{\hat{v}^{(r)}_s}{\hat{v}_s} \hat{e}_{(r)}+\f{\hat{v}^{(\p)}_s}{\hat{v}_s} \hat{e}_{(\p)}\right]  \, ,
    \label{eq:tetrad}
\eea
where $\hat{v}_s = \sqrt{\left(\hat{v}^{(r)}_s\right)^2+\left(\hat{v}^{(\p)}_s\right)^2}$.
For a ZAMO, $e_{[1]}$ and $e_{[3]}$ are found to be orthogonal and parallel to the fluid velocity, respectively.

As in \cite{Chen:2024ggq}, here we focus only on the fast reconnection model within the ideal MHD framework \cite{Liu2017}.  In this model, the thermal pressure is assumed to be negligible relative to the magnetic pressure, which generally exists in highly magnetized accretion systems, including magnetically arrested disks \cite{MAD,Tchekho2011}. Moreover, the magnetic field lines are assumed to have opposite directions upon crossing the equatorial plane and magnetic reconnection occurs within a much smaller region \cite{Priest1986} near the dominant reconnection $p$-point \cite{Titov1997} so that reconnection can be regarded to take place at the $p$-point on the global scale. Due to spontaneous perturbations, the two sets of anti-parallel magnetic field lines $\vec{B}_0$ curve toward each other, which which causes the local magnetic pressure to reduce and further push the anti-parallel magnetic lines towards each other more until they touch at the current sheet. Then reconnection occurs and its energy makes the plasmoid burst into two parts, ejecting along the direction of the upstream magnetic field line $\vec{B}_0$. In a word, the reconnection plays a role similar to that of ``bomb" in the Penrose process. In the fast reconnection model, the size of the current sheet can be comparable to the mean free paths of ions \cite{yamada2009}. The current sheet can be simply characterized by
the geometry index as
\bea\label{eq:geo_index}
    {\sf g}=\frac{d}{L} \, ,
\eea
where $d$ and $L$ denote the half-width and half-length of the current sheet, respectively. Assuming that the plasmoids do not slow down after leaving the current sheet and using the MHD equilibrium condition, the magnitude of their ejection velocities can be estimated as \cite{Chen:2024ggq}
\bea
    v_{out} \simeq \sqrt{\frac{\left(1-{\sf g}^2\right)\sigma_m}{1+\left(1-{\sf g}^2\right)\sigma_m}} \, ,
    \label{eq:vout}
\eea 
where $\sigma_m=B_{m}^2/\omega$ denotes the magnetization in the current sheet. Obviously, the ejection velocity tends to that of the Sweet-Parker model \cite{rela-SP} if ${\sf g}$ tends to zero. The limit $\lim\limits_{{\sf g}\to 1}v_{out}=0$ also imposes a restriction on the geometric index of the current sheet.

 The ratio $R_0$ of the local inflow speed to the Alfv{\'e}n velocity can be used to describe the generation rate of magnetic reconnection. When $\sigma\gg 1$, one has \cite{Chen:2024ggq}
\bea
    R_{0}\simeq {\sf g}\frac{1-{\sf g}^2}{1+{\sf g}^2}
                    \sqrt{\frac{\left(1-{\sf g}^2\right)^3 \sigma}{\left(1+{\sf g}^2\right)^2+\left(1-{\sf g}^2\right)^3 \sigma}} \, ,\label{eq:RHLM}
\eea
which has a peak-value at ${\sf g}\simeq 0.49$. As in \cite{Chen:2024ggq}, in the cases of high local-magnetization,  magnetic reconnection can be assumed to occur at the maximum local reconnection rate with the geometric index ${\sf g}=0.49$.

The magnetic field of a black hole is generally thought to be antiparallel adjacent to the equatorial plane, which is crucial for the occurrence of magnetic reconnection. For the plasma moving along non-circular plunging streamlines, we can assume the fluid is stationary, axisymmetric, and ideal-MHD because the resistivity of the plasma is nearly zero outside the current sheet so that the magnetic field lines are frozen in flowing plasma \cite{Ruffini:1975ne,Hou:2023bep}. For such a magnetic field, there exists a field line angular velocity $\Omega_B$ characterizing the rotation of magnetic field lines. For simplification, we here adopt the assumption of a zero field line angular velocity  $\Omega_B=0$ which  significantly reduces the complexity of calculations.
  From the equation $F^{\m\n}U_{\m} = 0$, one can find that for a stationary and axisymmetric fluid configuration composed of a zero-resistivity plasma in the Kerr-Taub-NUT spacetime, the global magnetic field can be expressed as 
\bea
    B_0^t = -\f{\Psi}{\sqrt{-g}} \bigg( U_r + \f{U^{\phi}}{U^r}L \bigg) , \quad\quad\quad B_0^r = -\f{\Psi}{\sqrt{-g}} E, \quad\quad\quad B_0^\phi = -\f{\Psi}{\sqrt{-g}} \f{U^{\phi}}{U^r} E ,
    \label{B}
\eea
where $\Psi$ denotes the overall strength and keeps constant along the streamline and  $g$ is the determinant of the metric. 
With the previous transformation (\ref{eq:tetrad}), one can obtain the magnetic field components in the fluid's rest frame \cite{Hou:2023bep}, i.e.,
\bea
    B_0^{[1]} = \f{\Psi}{\sqrt{-g} \,\hat{U}} \sqrt{g_{rr}g_{\p\p}}\,\o^{\p}\, , \quad \quad \quad B_0^{[3]} =  \f{\Psi}{\sqrt{-g} \,\hat{U} U^r}  (\hat{\g}_s E-\a) \,,
    \label{B1}
\eea
where $\hat{U} = \sqrt{\left(\hat{U}^{(r)}\right)^2+\left(\hat{U}^{(\p)}\right)^2}$. In the fluid's rest frame,  the plasmoids are ejected along the magnetic field lines. Therefore, the ejection direction of plasmoids can be described by the angle between the magnetic field lines and $e_{[3]} $,
\bea
    \xi  = \arctan{ \f{B_0^{[1]}}{B_0^{[3]}} } = \arctan{\bigg(\f{\sqrt{g_{rr}g_{\phi\phi}}\,\o^{\phi}U^r}{\hat{ \g}_s E -\a} \bigg)},. \label{angle}
\eea
For the fluid accreted by the black hole, it is easy to find that the angle $\xi$ is always negative since 
$U^r < 0$ and $\hat{ \g}_s \ge 1$. 

In the fluid's rest frame, the four-velocities of the ejected plasmoids can be expressed as
\bea
    u^{\m}\pl_{\m} =  \g_{out}\big[ e^{[0]}  \pm v_{out} \big(\cos{\xi} e^{[3]} + \sin{\xi} e^{[1]} \big)\, \big] \, ,
    \label{4v}
\eea 
where $\g_{out}$ is the Lorentz factor measured in the fluid's rest frame and the sign ``$\pm$” denotes two plasmoids towards opposite directions. In the reconnection process, one can neglect the magnetic energy because in the current sheet most of it is converted to the plasma's kinetic energy.
Furthermore, with the incompressible sphere approximation, one can find that the proper enthalpy density in the plasmoids is equal to that of the bulk plasma at the $p$-point \cite{Koide:2008xr,Chen:2024ggq}  and the enthalpy is $\D H = \o \D V$, where $\D V$ is a small proper volume of the plasmoids.
The energy per enthalpy at infinity for ejected plasmoids is  
\bea
 \ep = \f{\D E}{\D H} = \f{e \D \hat{V}}{\o \D V} = \f{e}{\a u^t \o} = -u_t - \f{\td{p}}{u^t},, 
\eea 
where $\D \hat{V} =  \D V/\hat{u}^{(t)} =   \D V/( \a u^t)$ is the volume of the plasmoid measured by the normal tetrad and $\D E = e \D \hat{V}$ is the energy. 
The quantity $\td{p} = p/\o$ is the pressure per enthalpy. The energy per enthalpy $ \ep$ represents the magnitude of energy obtained by the plasmoid from magnetic
reconnection. Inserting $u^t$ and $u_t$ into the above equation, one has  \cite{Chen:2024ggq} 
\begin{align}\label{ep}
\ep_{\pm} = &\, \, \a \hat{\g}_s \g_{out} \bigg[  \big(1+\b^{\p}\hat{v}^{(\p)}_s\big) \pm  v_{out}\bigg(\hat{v}_s+ \b^{\p}\f{\hat{v}^{(\p)}_s}{\hat{v}_s}\bigg)\cos{\xi} \mp v_{out}\b^{\p}\f{\hat{v}^{(r)}_s}{\hat{\g}_s \hat{v}_s} \sin{\xi}   \bigg] \nn \\
&- \f{\a \, \td{p}}{\hat{\g}_s \g_{out}\big(1\pm \hat{v}_s v_{out} \cos{\xi}\big)} \, .
\end{align}
 The spacetime  affects the energy per unit enthalpy of the plasmoid not only directly through the factors $\a$ and $\b^{\p}$, but also indirectly through the fluid velocity and the magnetic field angle.

\section{Energy extraction by magnetic reconnections in the Kerr-Taub-NUT spacetime}
\label{sec:EE_inspiral}

Let us now investigate energy extraction in the plunging region from a Kerr-Taub-NUT black hole by magnetic reconnections. Here we assume that the plasma plunges into the black hole from ISCO under certain perturbations and only consider the case where the plasma moves along the prograde direction of the black hole. The radial velocity of the plunging geodesics can be obtained from $U_\m U^\m=-1$.
\begin{figure}[h!]
	\centering
    {\includegraphics[width=0.48\textwidth]{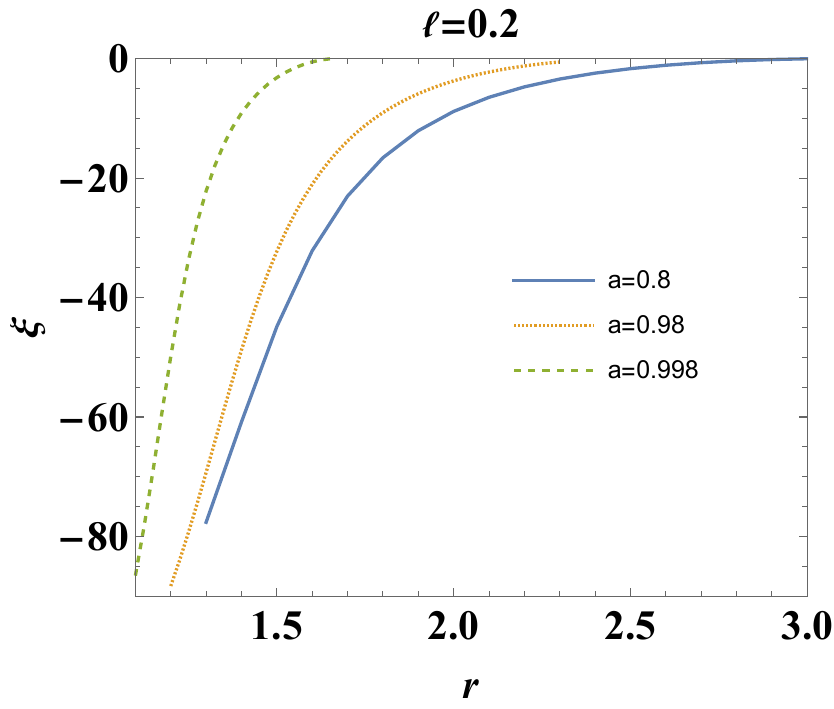}}\,
	{\includegraphics[width=0.48\textwidth]{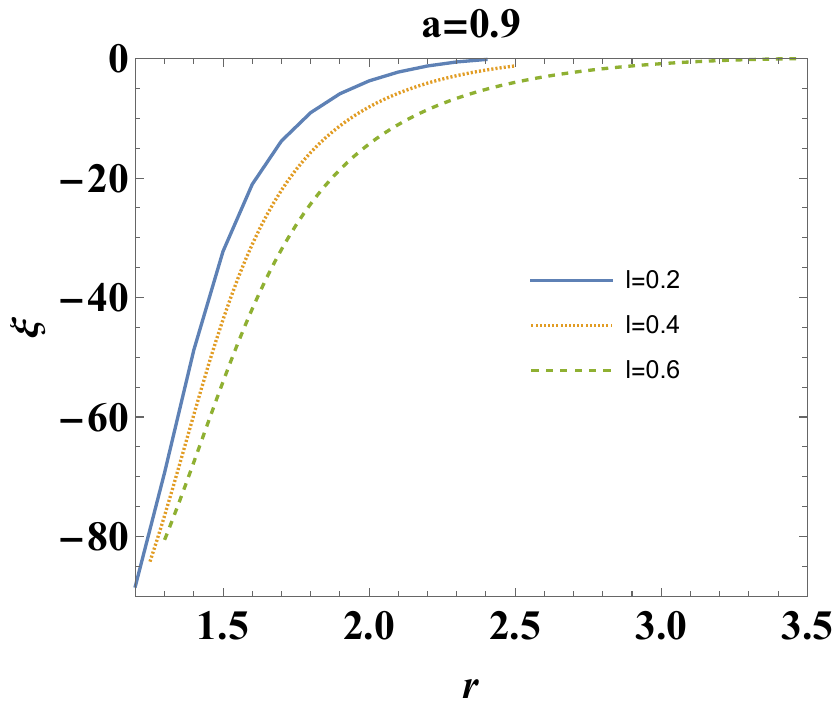}}\,\\
    {\includegraphics[width=0.48\textwidth]{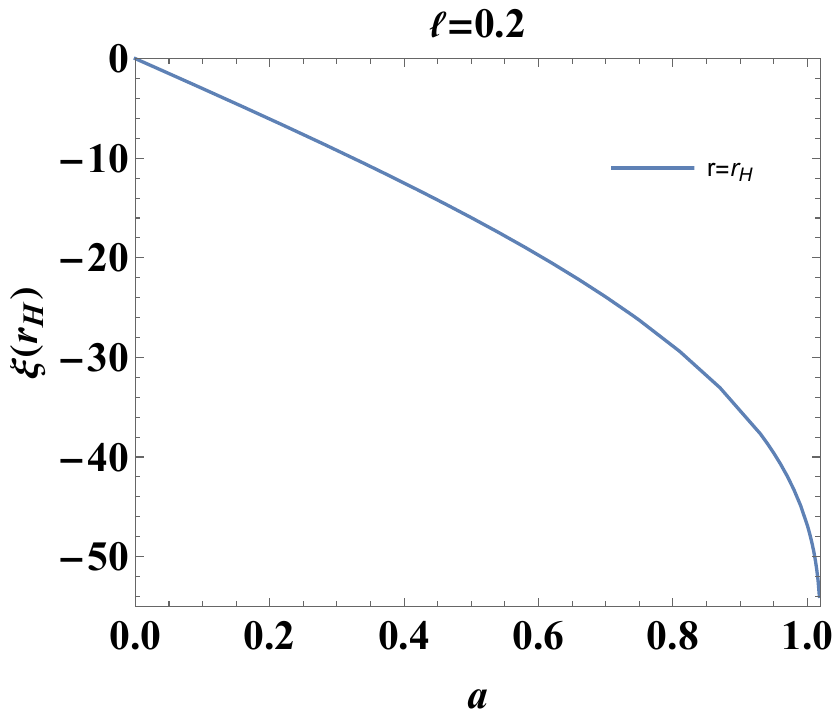}}\,
	{\includegraphics[width=0.48\textwidth]{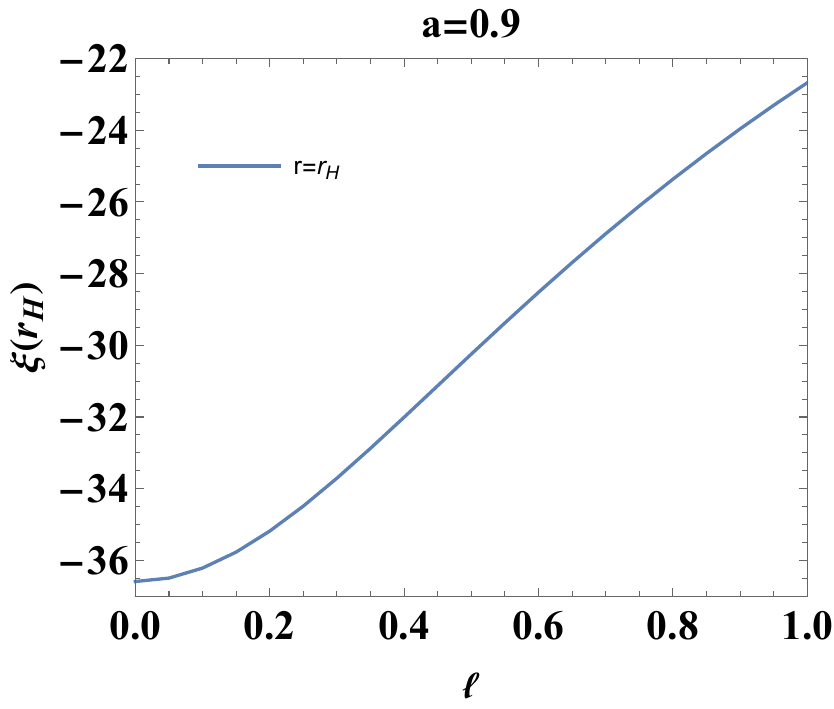}}
	\caption{ Behaviors of the magnetic field angle in the bulk plasma plunging from the ISCO in Kerr-Taub-NUT geometry. Top: Change of the magnetic field angle with the radial coordinate $r$ for different black hole spin $a$ and gravitomagnetic charge $l$. Bottom: The value of $\xi$  at the event horizon $r_H=M+\sqrt{M^2-a^2+l^2}$  with different $a$ and $l$. }
	\label{xi}
\end{figure}

In Fig.~\ref{xi} we present effects of the black hole spin $a$ and gravitomagnetic charge $l$ on the magnetic field angle. In the Kerr-Taub-NUT spacetime, the angle $\xi$ for the plunging plasmoid is also negative and its absolute value increases as the plasmoid gets closer to the black hole's horizon, which is similar to those of the Kerr case. The presence of gravitomagnetic charge $l$ decreases the change rate of the magnetic field angle $\xi$ with respect to $r$. The absolute value of $\xi$  at the horizon decreases with the gravitomagnetic charge $l$. In the case $l=0.2$, the magnetic field angle at the horizon approaches a minimum value of $-54^{\circ}$ as $a \rightarrow 1.019$, whose absolute value is less than that in the Kerr case where it tends to $-60^{\circ}$ as $a\rightarrow 1$.
These results show that  effects of the gravitomagnetic charge on the angle $\xi$ are opposite to those of the black hole spin.

\begin{figure}[h]
	\centering
    {\includegraphics[width=0.32\textwidth]{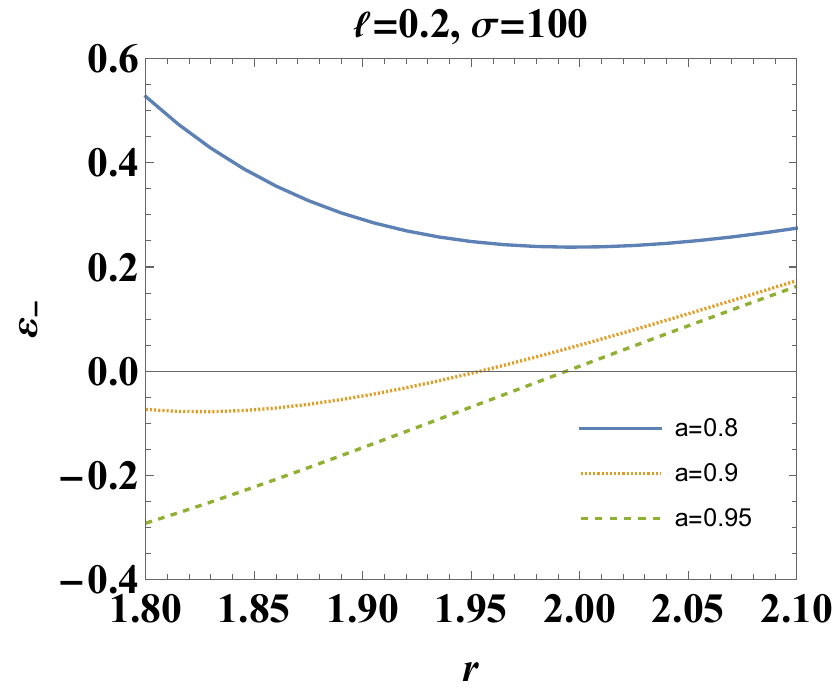}}\,
	{\includegraphics[width=0.32\textwidth]{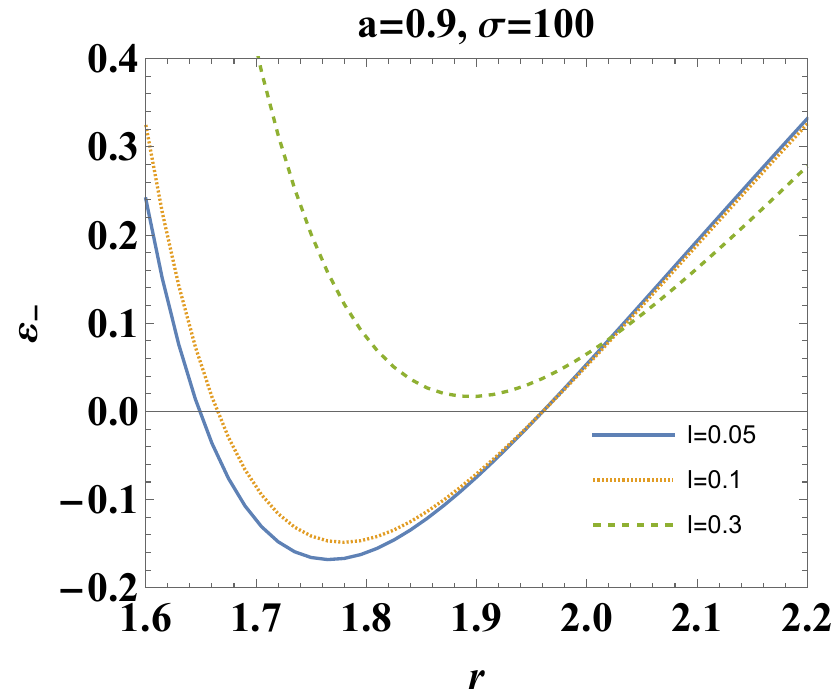}} \,
	{\includegraphics[width=0.32\textwidth]{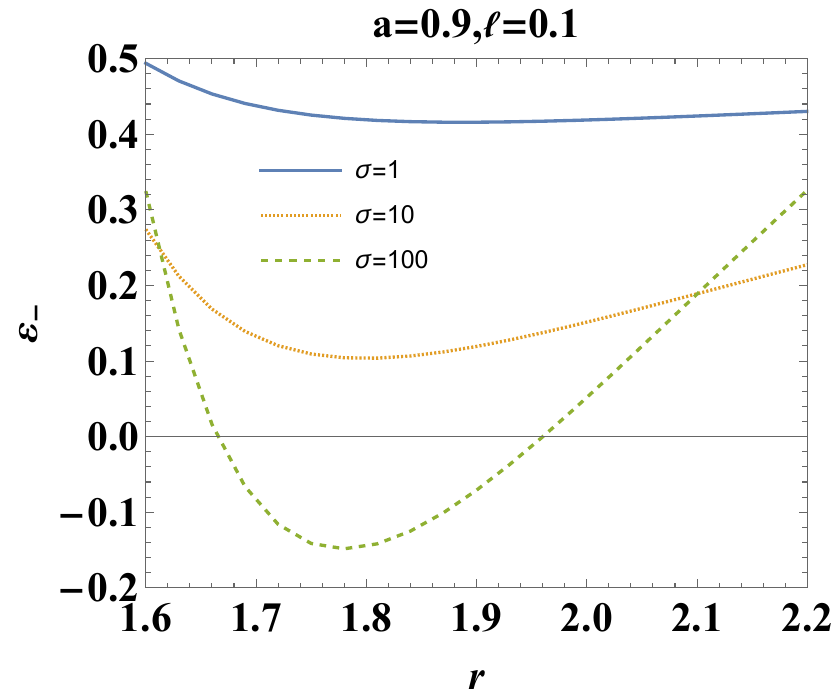}}
	\caption{The energy per unit enthalpy $\epsilon_-$ at infinity of the ejected plasmoids varies with the reconnection location, for different values of $a$, $l$ and $\sigma$ in $\td{p} = 1/4$.}
	\label{epins}
\end{figure}

\begin{figure}[h!]
	\centering
	{\includegraphics[width=0.48\textwidth]{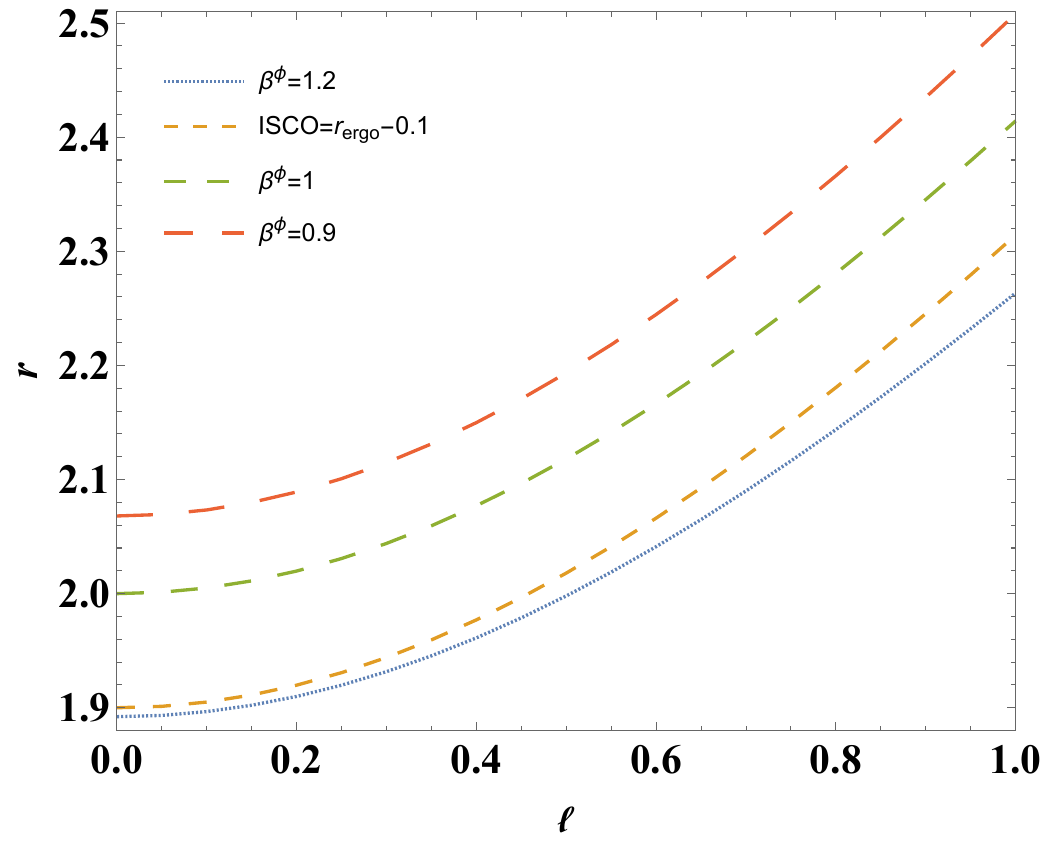}}
	\caption{The radii of reconnection locations with the gravitomagnetic charge $l$ for different value of $\beta^\phi$.}
	\label{beta1}
\end{figure}

To extract energy from the black hole through magnetic reconnection, it
is anticipated that the decelerated plasma should exhibit negative energy as observed at infinity, while the accelerated plasma is expected to have positive energy. Thus, in Fig.~\ref{epins}, we present the changes in the energy per enthalpy $\epsilon_{-}$ of the ejected plasmoids measured at infinity with the locations of reconnection occurred.  For simplicity, we neglect the variation of the bulk plasma magnetization $\s_0$ with the coordinate $r$. In general, a smaller gravitomagnetic charge $l$, a larger black hole spin parameter $a$, and a higher magnetization $\sigma $ collectively facilitate the generation of negative energy$\epsilon_-$ and enhance the efficiency of energy extraction. Moreover, one can see that $\epsilon_-$ is not negative as the reconnection location is near the ISCO. As the radius of the location of the reconnection decreases, the negative $\epsilon_-$ gradually emerges and reaches its minimum value. Finally,  it vanishes rapidly as it approaches the black hole event horizon. The radial coordinate at which the minimum value of $\ep_-$ appears decreases with black hole spin $a$ and increases with the gravitomagnetic charge $l$. Moreover, the radius of the reconnection location increases with the gravitomagnetic charge $l$ for fixed $\beta^{\phi}$, which is shown in Fig.~\ref{beta1}.

 If the plasma particles with negative energy $\ep_{-}$ fall past the outer event horizon into the black hole along the corotating direction, while the plasma particles with positive energy $\ep_{+}$ escape to infinity along the counter-rotating direction and carry away more energy, the efficiency of energy extraction via magnetic reconnection can be defined as \cite{Chen:2024ggq} 
\be\label{etad}
\eta = \f{\ep_{+}}{\ep_{+}+\ep_{-}} \, .
\ee
Obviously, only negative $\ep_-$ permits $\eta>1$, which is satisfied for the plunging plasma.
\begin{figure}[ht]
	\centering
    {\includegraphics[width=0.32\textwidth]{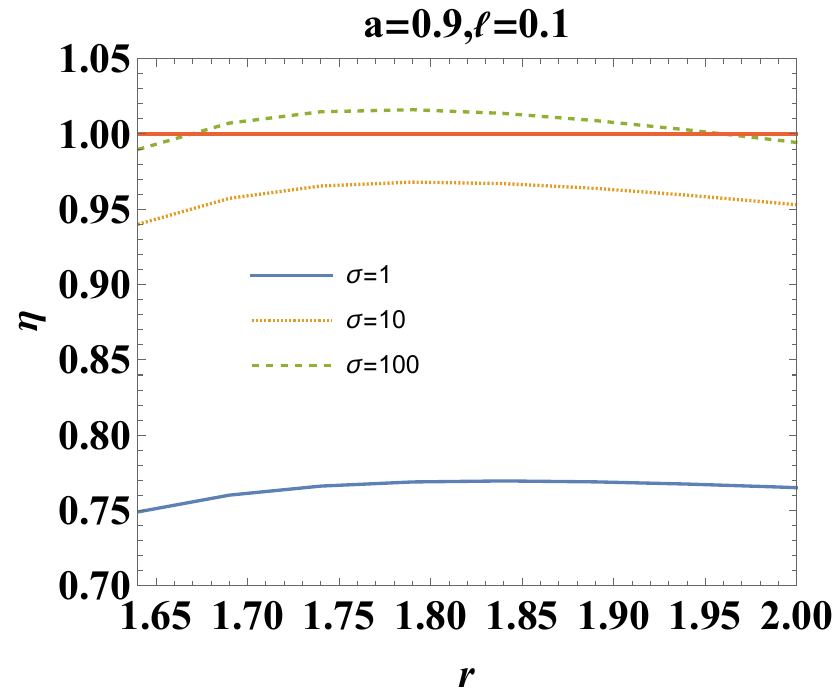}}\, 
	{\includegraphics[width=0.32\textwidth]{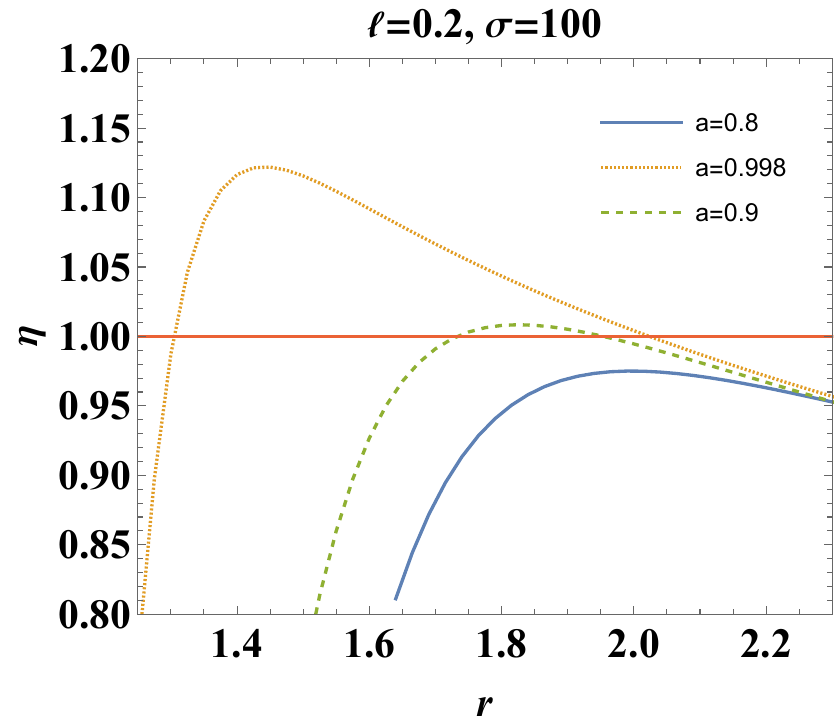}} \,
	{\includegraphics[width=0.32\textwidth]{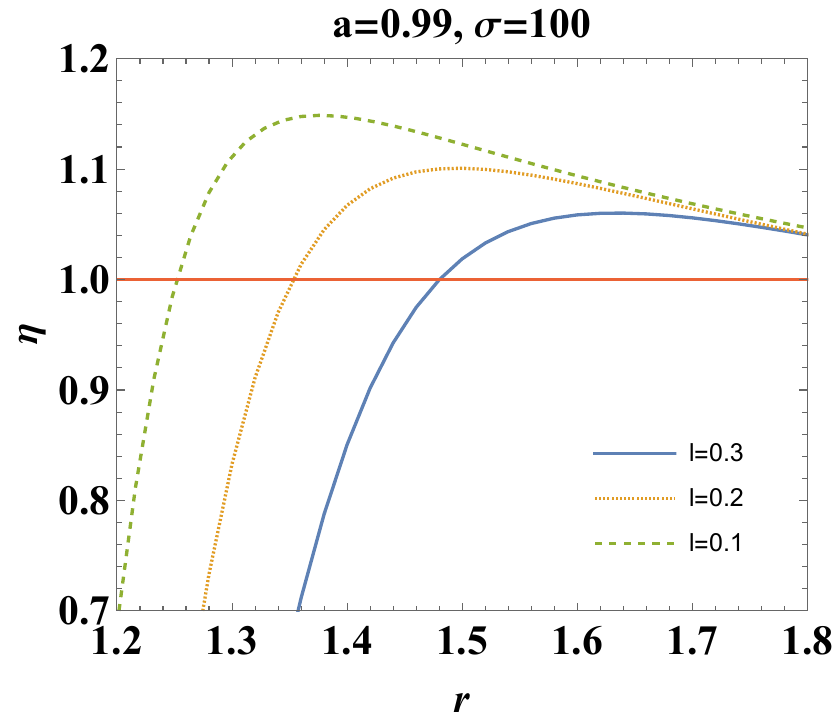}}
	\caption{The energy extraction efficiency as a function of the reconnection location for different values of $\s$ (Left),  for different values of $a$ (Middle) and  for different values of $l$ (Right).}
	\label{eta}
\end{figure}
\begin{figure}[ht]
	\centering
	{\includegraphics[width=0.40\textwidth]{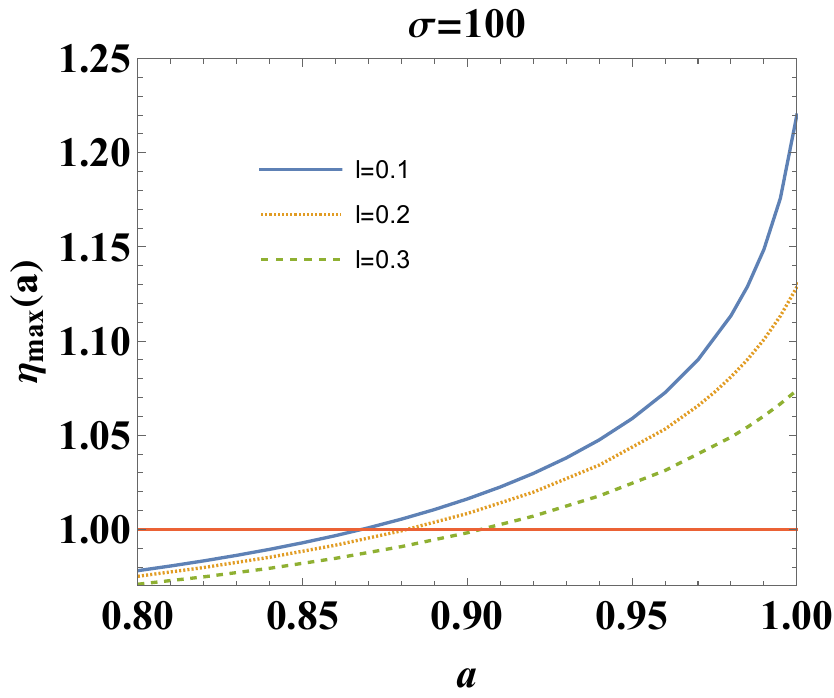}}\, 
	{\includegraphics[width=0.40\textwidth]{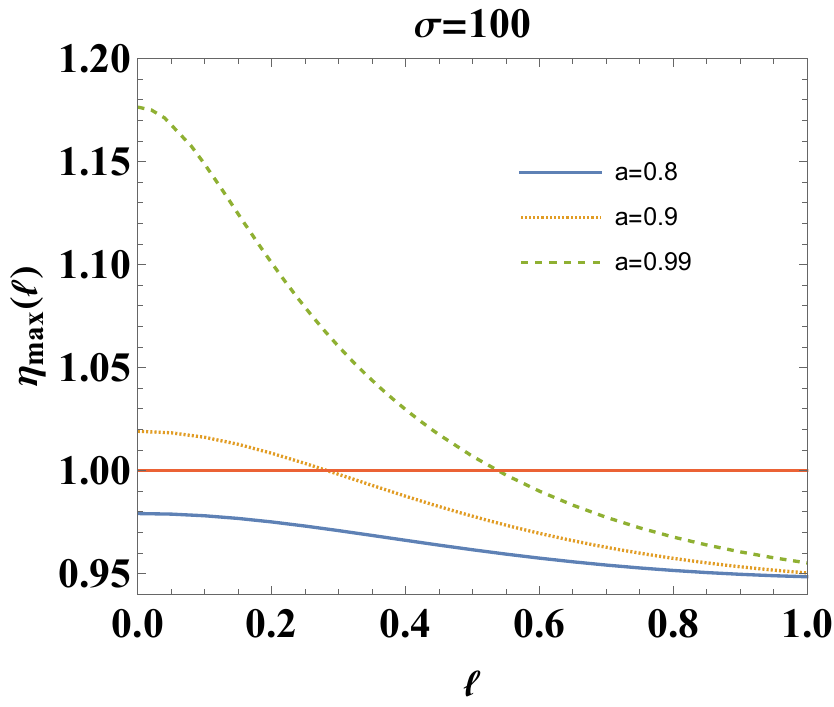}} \,\\
	{\includegraphics[width=0.40\textwidth]{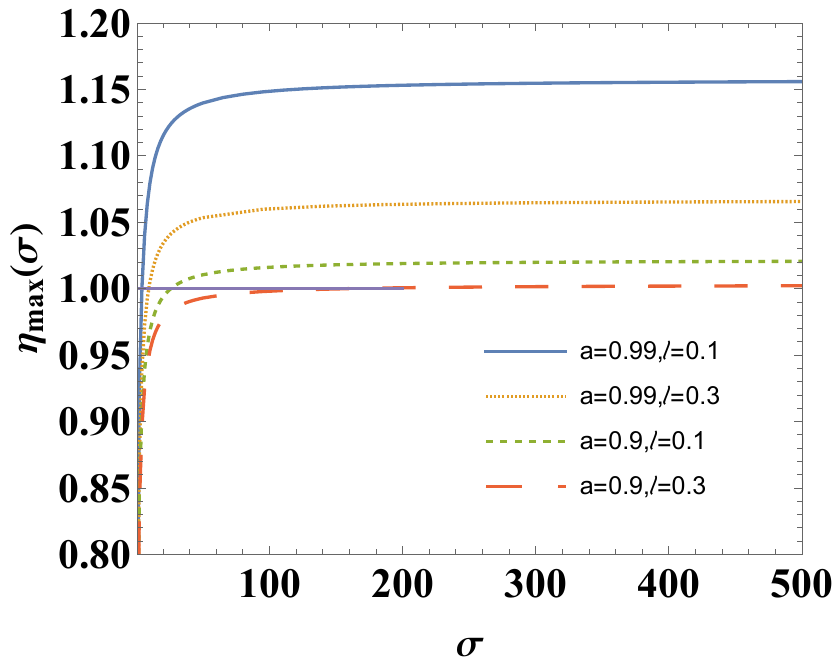}}\.
    {\includegraphics[width=0.40\textwidth]{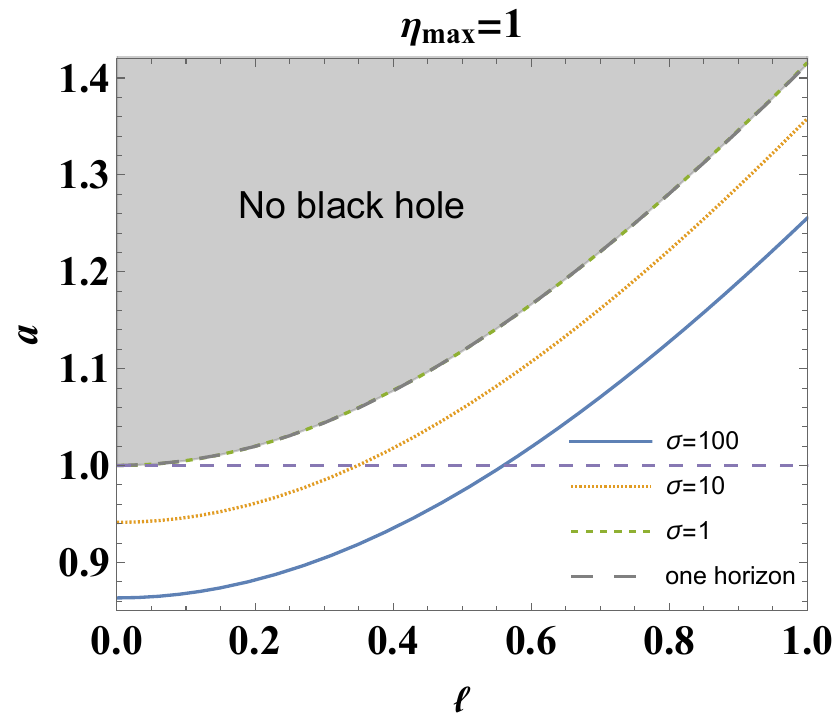}}
	\caption{The maximal value of $\eta$ as a function of $a$, $l$ and $\sigma$, respectively. Bottom right: the relationship curves between $a$ and $l$ when the maximal value $\eta_{max}=1$ for different magnetization.}
	\label{eta1}
\end{figure}
In Fig.~\ref{eta}, we present the dependence of the local energy extraction efficiency on the location of reconnection within the ISCO. The extraction efficiency increases with the magnetization parameter $\sigma$ and the black hole spin parameter $a$, which is similar to those in the Kerr black hole case. However, it decreases with the gravitomagnetic charge $l$. For a strong magnetization scenario, we also find that the energy extraction is also located near the ISCO in the Kerr-Taub-NUT black hole spacetime.

 Fig.~\ref{eta1} displays the maximum value of $\eta$ under various spins, gravitomagnetic charges and magnetization parameters. We find that the range of $a$ at which the energy extraction efficiency exceeds $1$ decreases with the gravitomagnetic parameter $l$. In other words, in order to attain an equivalent peak efficiency, the Kerr–Taub-NUT black hole demands a spin significantly higher than its Kerr counterpart. Moreover, the maximum value $\eta_{max}$ increases with the magnetization parameter $\sigma$ and almost reaches a constant once $\sigma$ exceeds approximately $100$. This means that one can use $\sigma = 100$ to represent cases of extremely high magnetization as in the Kerr case \cite{Chen:2024ggq}.
 In the bottom right panel of Fig.~\ref{eta1}, we plot the relationship curves between spin and gravitomagnetic charge under the condition that the maximum value of $\eta=1$. In the parameter plane $l-a$, the effective energy extraction  (that is, $\eta>1$) occurs only in the region between the curve corresponding to $\eta_{max}=1$ and the one-horizon curve (that is, $a^2-l^2=M^2$). It is notable that for low magnetization $\sigma\simeq1$, there is nearly no allowed parameter region because the curve $\eta_{max}=1$ overlaps with the boundary without black hole existence. 
 
We now study the condition for the plasmoid with energy $\ep_+$ to escape to infinity, which is essential for astronomical observations. In the regime of strong magnetization, we can treat plasmoid as a timelike particle. One of reasons is that the plasmoid after leaving the current sheet possesses a relativistic ejection velocity and nearly moves free \cite{Priest1986, Titov1997}.  
Moreover, the thermal energy in the plasmoid radiates away and do not converted into the kinetic energy of the particle, which means that it does not affect the plasmoid's trajectory.
The radial motion of the plasmoid's in the spacetime can be described by equation
\begin{eqnarray}
g_{rr} u_r^2 + g_{\theta\theta} u_\theta^2 = V_{\rm eff} ,
\end{eqnarray}
with $u_\theta=0$. The effective potential $V_{\rm eff}$ is
\begin{eqnarray}
V_{\rm eff}=\frac{E^2 g_{\phi \phi}+2 E L g_{t \phi}+L^2 g_{tt}}{g_{t\phi}^2-g_{tt}g_{\phi \phi}}-1.
\end{eqnarray}
Using the conserved energy $E$ and angular momentum $L$ from the ejection velocity Eq.~\eqref{4v}, we can obtain the escaping conditions. 

\begin{figure*}[!h]
	\centering
    {\includegraphics[width=0.4\textwidth]{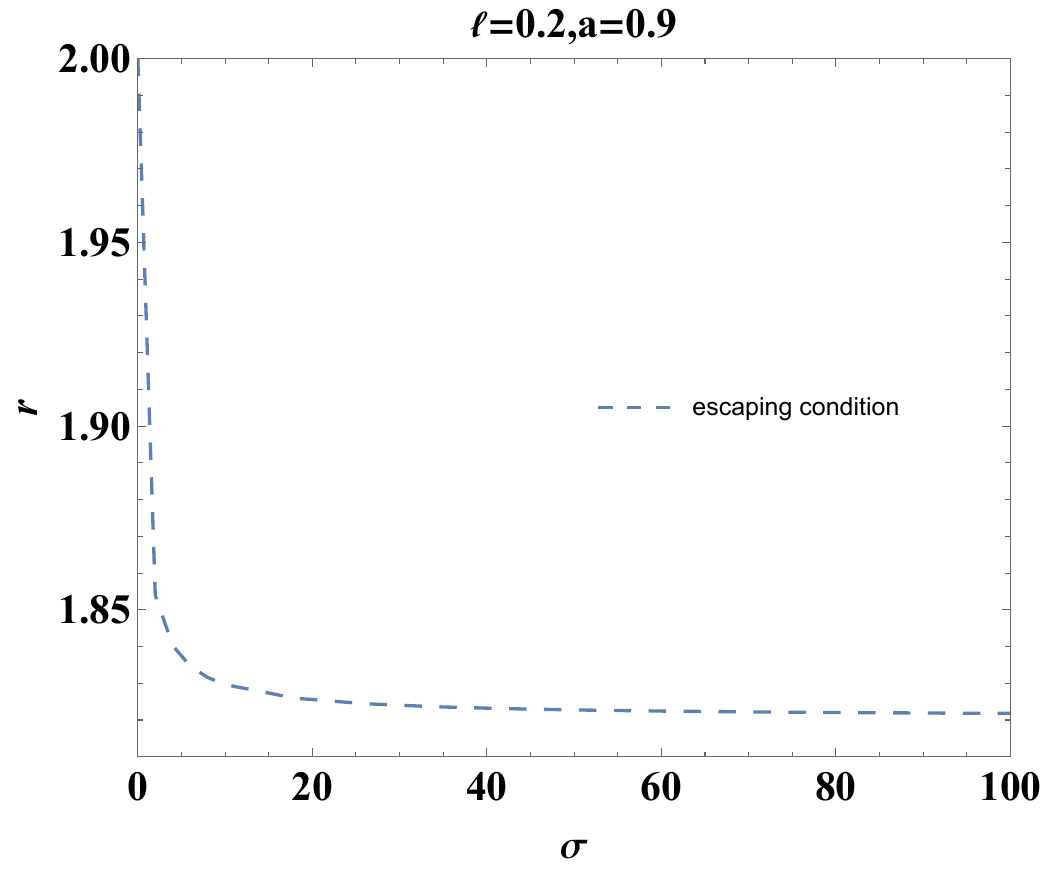}}\,
    {\includegraphics[width=0.4\textwidth]{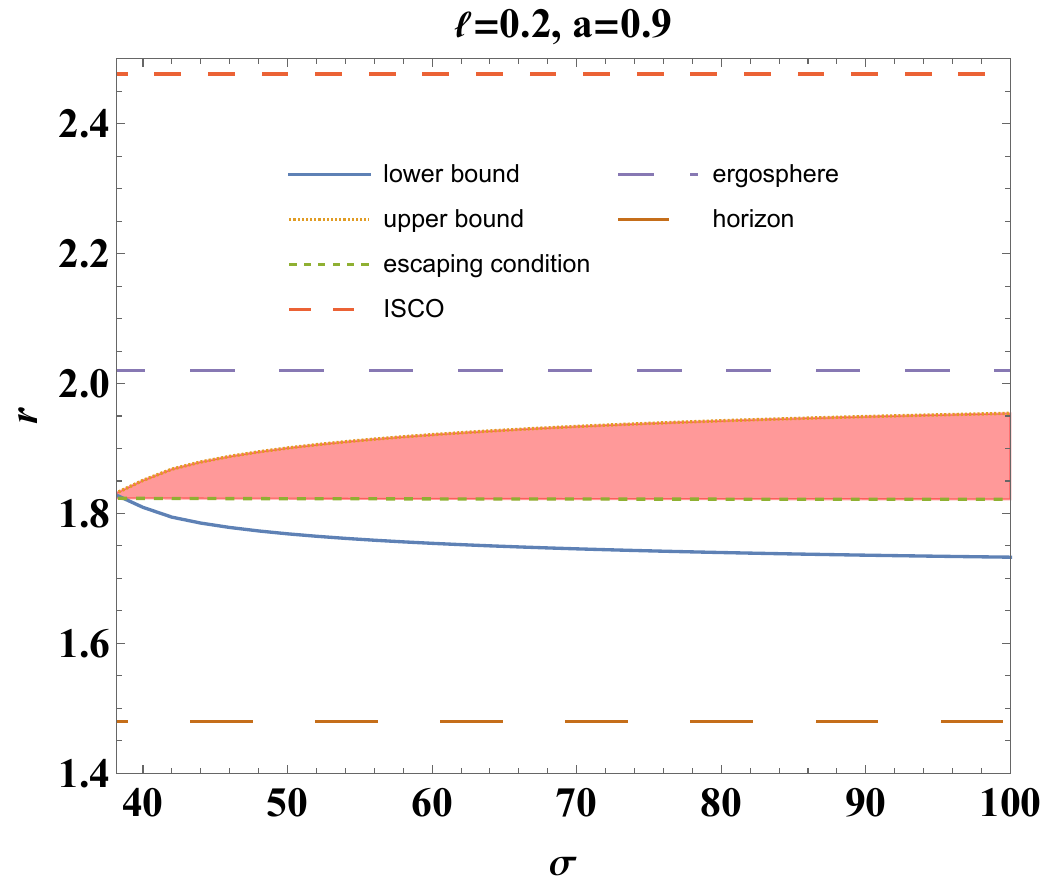}}\,\\
	{\includegraphics[width=0.4\textwidth]{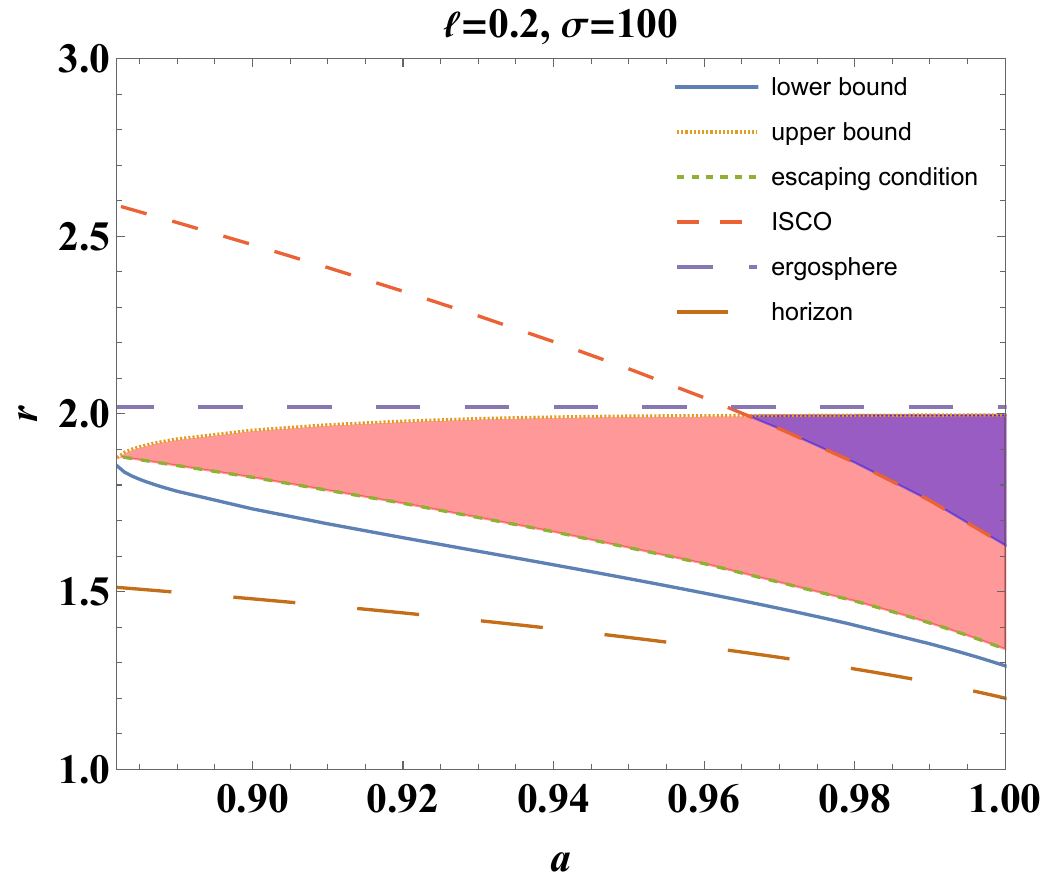}}\,
	{\includegraphics[width=0.4\textwidth]{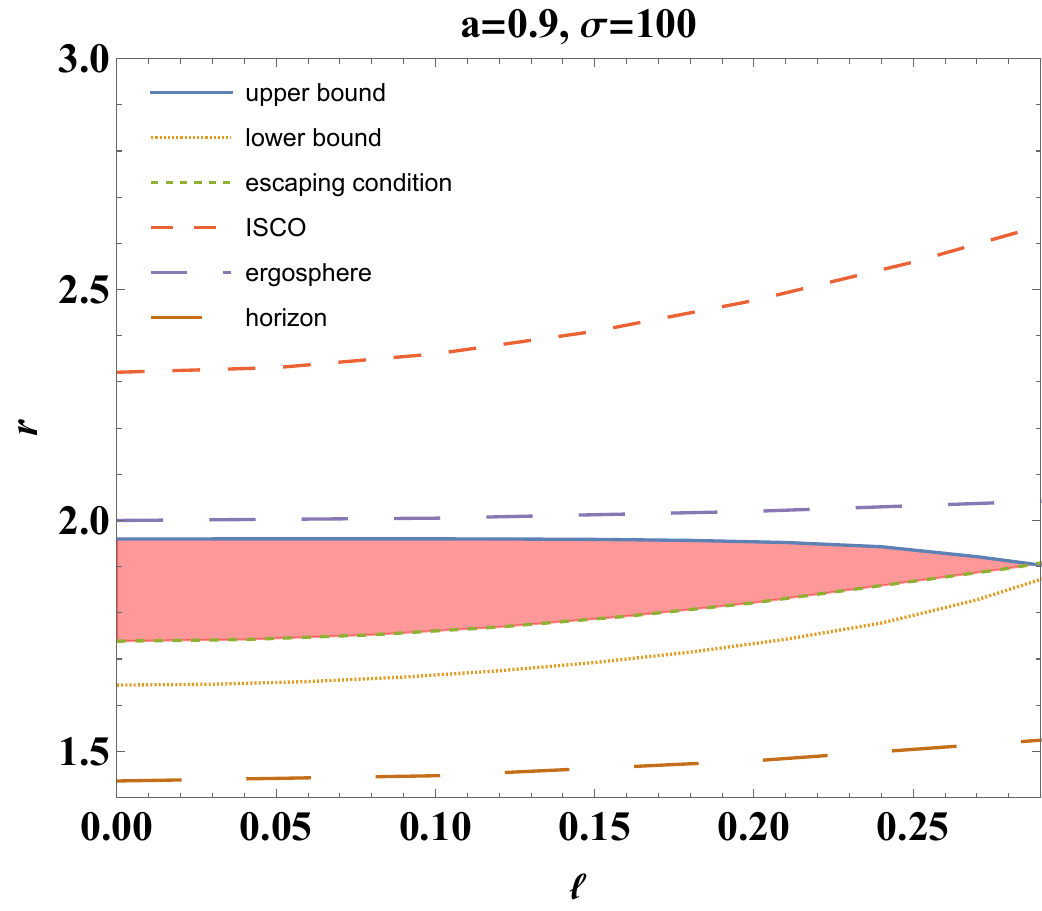}}
	\caption{Changes of the critical radius with black hole spin $a$, magnetization parameter $\sigma$ and NUT charge $l$.   The region in red and purple means extracting energy can escape to the infinity and extraction efficiency $\eta$ over $1$. Only between the orange and blue dashed curves can $\eta$ be above 1. The green dashed curve stands for the critical radius for the escaping condition, above which the plasmoid with $\ep_+$ can escape to the infinity.}
	\label{escape2}
\end{figure*}

From Eqs.~\eqref{angle} and ~\eqref{4v}, one can find that for constant $\sigma$ and $\sf g$ (hence a constant $ v_{out}$) $E$ and $L$ can be rewritten in the forms as
\bea
E = E \left(\, r_0,\, \xi(r_0) \,\right) \, , \quad  L = L \left(\, r_0,\, \xi(r_0) \,\right) \, , \label{elll}
\eea
where $r_0$ denotes the radial coordinate of the $P$-point at which the magnetic reconnection occurs. From the previous analysis, it is easy to find that the radial velocity of the plasmoid with $\ep_+$ always points inward. To ensure that the plasmoid with $\ep_+$ can escape to infinity,  the potential at $r = r_0$ must be large enough to reverse the direction of the plasmoid and propel it outward. In general, the potential has an extreme at the point $r=r_c$, where $r_c$ can be written as a function of the radius of the $P$-point in terms of Eq.\ref{elll}, i.e., $r_c = r_c(r_0)$. Therefore, the escape condition of the plasmoid with $\ep_+$ can be expressed as 
\bea\label{con}
V\left(\,r_c(r_0),\,E(r_0),\,L(r_0)\,\right)<0  \, , \quad \, \,  \text{if} \quad  r_H<r_c(r_0) < r_0 \, 
\eea	
Where $V=V_{\rm eff}/g_{rr}$. As $r_c(r_0) > r_0$, the plasmoid does not escape because it cannot be bounced outward and consequently falls into the horizon \cite{Chen:2024ggq}. The condition ~\eqref{con} gives a critical radius $r = r_c$ and only in the case $r_c(r_0) < r_0$ the plasmoid is allowed to escape.
Fig.~\ref{escape2} presents changes of the critical radius with black hole spin $a$, magnetization parameter $\sigma$ and NUT charge $l$.
It shows that the critical radius $r_c$ increases with the NUT charge $l$. Moreover, we find that the critical radius decreases
with the magnetization parameter. Especially, the decreasing rate slows rapidly with increasing magnetization. It should be noted that the variation of the escape condition in the radius-magnetization plane is opposite to that of $\eta_{max}$. It is understandable because the larger magnetization yields the larger $\ep_+$. We also illustrate the permitted region (marked in red) of effective energy extraction for the plunging plasma in the radius-magnetization, radius-spin and radius-gravitomagnetic charge planes. Magnetic reconnection occurring in this region results in energy extraction with efficiency exceeding 1 and the plasmoids can transfer the black hole's rotational energy to infinity. From Fig.~\ref{escape2}, we find that the region for effective energy extractions decreases with the gravitomagnetic charge $l$ and increases with the black hole spin. The purple region in the bottom left panel of Fig.~\ref{escape2} corresponds to the case where the ISCO is located outside the ergosphere when the black hole spin is rather large, and plasma plunges from the ergosphere instead of the ISCO.

\section{Explanation to jet power of source GRS1915+105}
\label{sec:ExpGRS1915}

It is of significance to constrain black hole parameters because it not only facilitates the unambiguous identification of black holes, but also provides a stringent test for gravitational theories. Recently, the jet power and the radiative efficiency of disks have been applied to constrain black hole parameters. However, the Kerr-Taub-NUT metric is found to not simultaneously explain the observed jet power and radiative efficiency of GRS 1915+105 \cite{PhysRevD.108.103013}. Here, we try to use the energy extraction mechanism via magnetic reconnection to revisit this problem.

\begin{figure*}[!h]
	\centering
    {\includegraphics[width=0.6\textwidth]{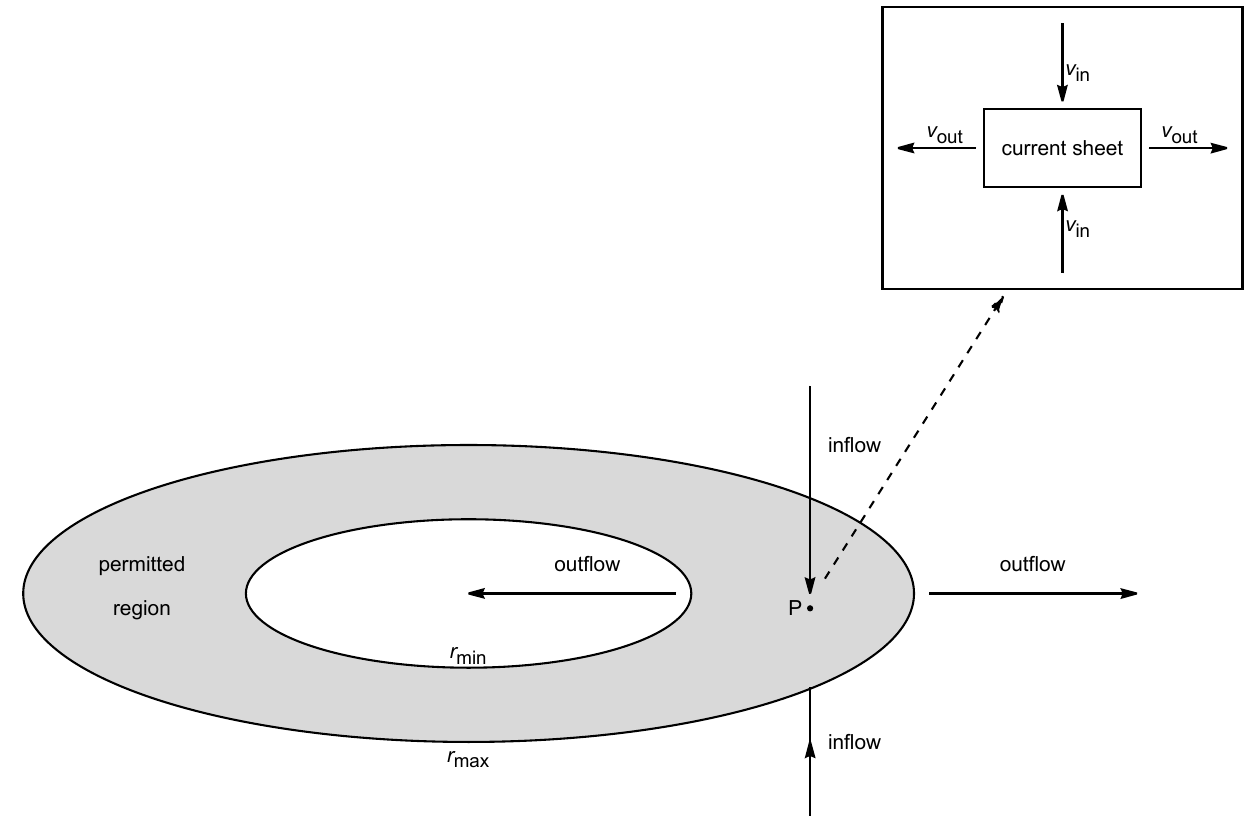}}
	\caption{The sketch of local magnetic reconnection occurred near the point $P$ and the globe flows.}
	\label{totalpower}
\end{figure*}

\begin{figure*}[!h]
	\centering
        {\includegraphics[width=0.4\textwidth]{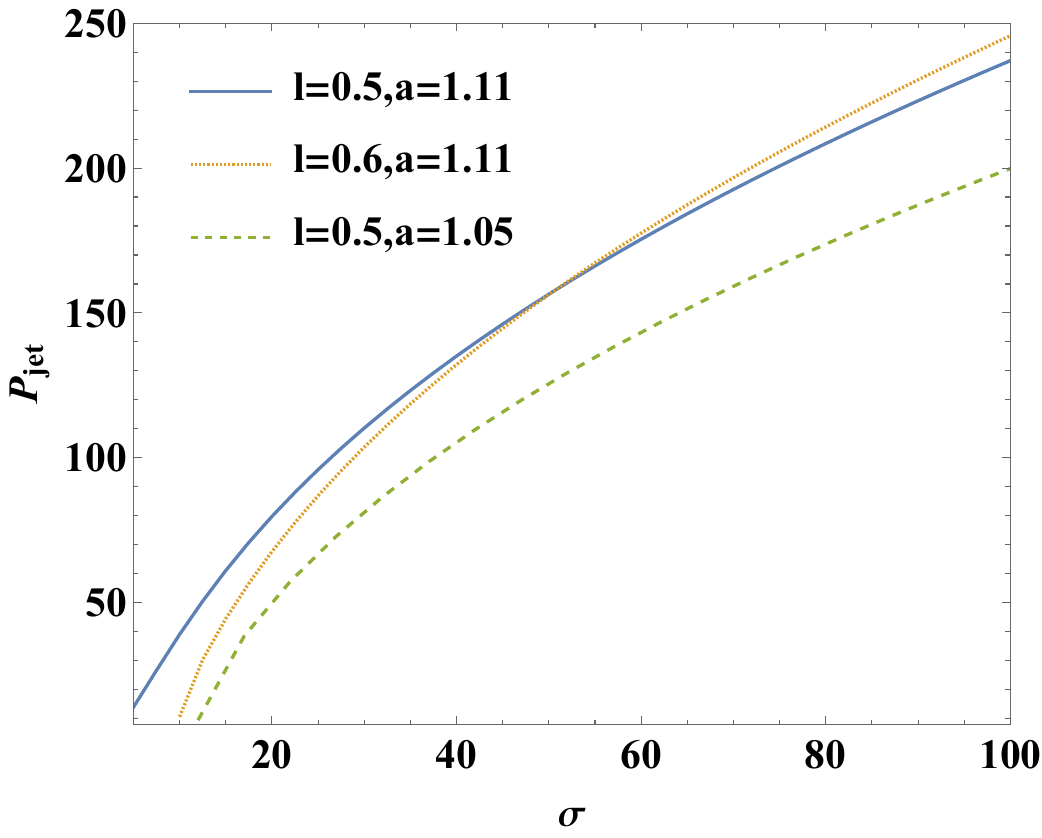}}
	\caption{Change of the total power $P_{jet}$  with magnetization $\sigma$ for different spin $a$ and gravitomagnetic charge $l$ values.}
	\label{jet}
\end{figure*}

In terms of the mass conservation in magnetic reconnections, we have $v_{in}L=v_{out}d$ in the current sheet and $j_{inflow}=j_{outflow}$ in the equatorial plane as shown in Fig.~\ref{totalpower}. Moreover, according to the distribution of the stream lines of fluid around the black hole together with the ideal MHD condition, one can obtain that the magnetic field lines converge into the equatorial plane, which means that the magnetic reconnection can occur in the wider region with the higher rate. 
 Setting $d=0.49L$, we have $v_{in}=0.49v_{out}$ and find that the total flux of the ejected plasmoids can be expressed as 
  \bea
  j=\int_{r_{min}}^{r_{max}}dr\sqrt{-g}1.96\pi rv_{out}(r),
  \eea
and 
the total jet power is 
\bea
P_{jet}=\int_{r_{min}}^{r_{max}}dr\sqrt{-g}0.98\pi rv_{out}(r)\epsilon_+(r)\,,
\eea
where plasmoids are assumed to ejected in opposite directions with equal amounts. In Fig. \ref{jet}, we present the changes of the total power $P_{jet}$  with the magnetization parameter $\sigma$ for different spin $a$ and gravitomagnetic charge $l$ values, which indicates that  the total power $P_{jet}$ increases with the gravitomagnetic charge $l$ in the higher magnetization parameter $\sigma$ case and decreases with in the lower $\sigma$ case.
\begin{figure*}[!h]
	\centering
    {\includegraphics[width=0.4\textwidth]{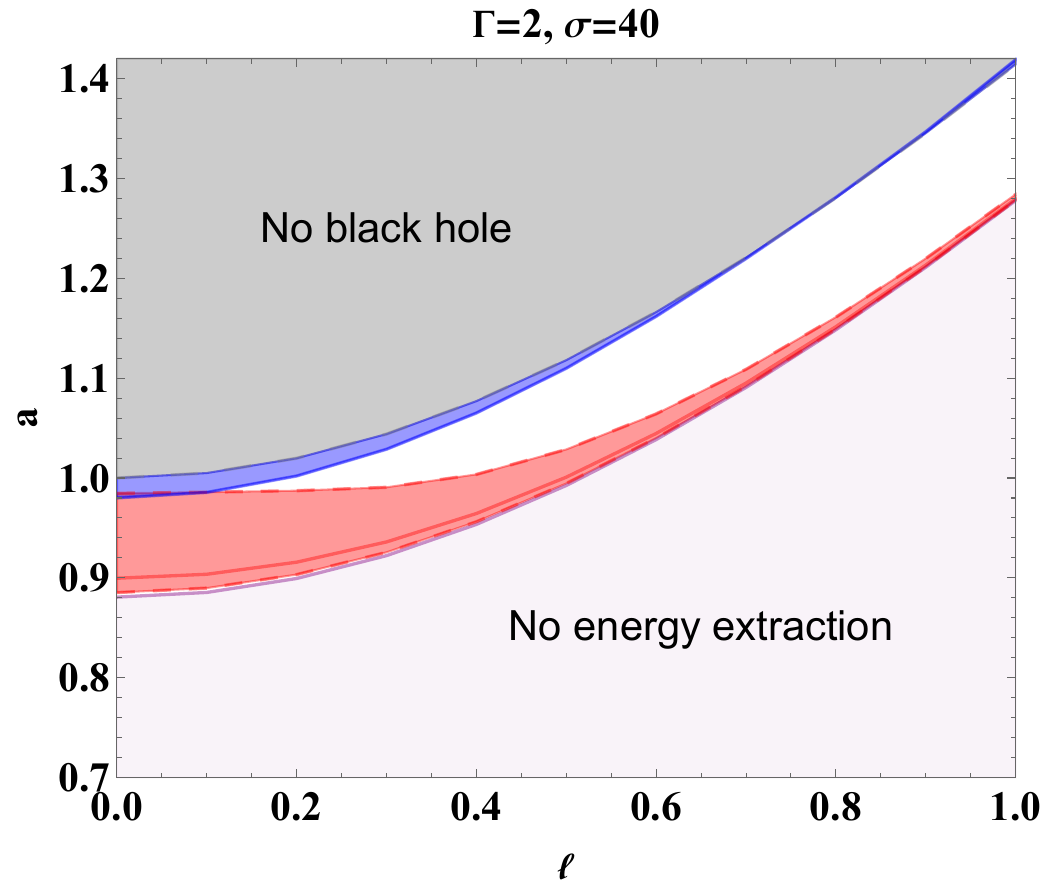}}
\caption{ The allowed regions in the $l-a$ plane for the Kerr-NUT black hole. The red colored region is obtained from the observed jet power of GRS1915+105 by the magnetic reconnection mechanism with the Lorentz factor $\Gamma=2$ and magnetization $\sigma=40$. The blue region is obtained by the observed radiative efficiency for GRS1915+105 and the blue solid curve corresponds to the case with radiative efficiency $\eta=0.234$. The red solid line is for the central value of the jet power $P_{jet}=42$ while the red dashed lines correspond to error of $0.3 dex$.}
\label{power}
\end{figure*}

For the source GRS1915+105,  the total power is observed as $P_{jet}=42$ with the error of 0.3 $dex$ for $\Gamma=2$ and radiative efficiency is in the range $\eta>0.234$ from the data listed in \cite{PhysRevD.108.103013}. 
With these data, in Fig. \ref{power}, we present the allowed region in the parameter plane $l-a$. We find that for $\sigma=40$ the allowed parameter region originating from the jet power has an intersection with the region from the radiative efficiency. This implies that through energy extraction via magnetic reconnection mechanism the Kerr-Taub-NUT metric is found to simultaneously explain the observed jet power and radiative efficiency of GRS 1915+105, which is different from that obtained by the Blandford-Znajek mechanism.

\section{Summary}
\label{sec:sum}
 
We have studied the energy extraction from a Kerr-Taub-NUT black hole via magnetic reconnections occurring in the plunging region. For plunging plasmoids, the presence of gravitomagnetic charge $l$ decreases the absolute value of magnetic field angle $\xi$ at the horizon and its change rate with respect to $r$, which means that effects of the gravitomagnetic charge $l$ on the magnetic field angle are opposite to those of the black hole spin. Moreover, the gravitomagnetic charge $l$ is not conducive to the generation of negative energy $\epsilon_-$ and reduces the efficiency of energy extraction, while increasing the radial coordinate at which the minimum value of $\ep_-$ appears.  Furthermore, the gravitomagnetic parameter $l$ shortens and decreases the range of values for the black hole spin $a$ in which the energy extraction efficiency $\eta>1$, which means that to attain an equivalent peak efficiency, the Kerr–Taub-NUT black hole demands a spin significantly higher than its Kerr counterpart. We also investigated the critical radius $r_c$ obtained by the escape condition of the plasmoid with $\ep_+$ and found that it increases with the NUT charge $l$. Overall, the gravitomagnetic charge suppresses the energy extraction process through magnetic reconnection, which is opposite to those of the black hole spin and the magnetization parameter.

Finally, we treat the energy extraction process through magnetic reconnection as a mechanism to revisit the problem on the observed jet power and radiative efficiency of GRS 1915+105. Our results show that the allowed black hole parameter region originating from the jet power has an intersection with the region from the radiative efficiency for the GRS 1915+105. This means that the Kerr-Taub-NUT metric is found to simultaneously explain the observed jet power and radiative efficiency for GRS 1915+105, which is not explained in previous studies. It could help us to understand energy extractions around real black holes in the astronomical environment.

\bibliographystyle{utphys}
\bibliography{magentic_reconnection_Kerr-nut}
		
\end{document}